\documentclass{aastex}
\usepackage{emulateapj5}
\usepackage{epsf,graphics}
\shorttitle{Clustering of Halos on the Light-cone}
\shortauthors{Hamana, Yoshida, Suto, \& Evrard}

\begin{document}

\title{Clustering of dark matter halos on the light-cone:\\ scale-,
    time- and mass-dependence of the halo biasing \\ in the Hubble
    volume simulations}

\author{Takashi Hamana\altaffilmark{1,2},  
Naoki Yoshida\altaffilmark{2}, Yasushi Suto\altaffilmark{3} 
and August E. Evrard\altaffilmark{4} 
}

\email{hamana@yukawa.kyoto-u.ac.jp, naoki@mpa-garching.mpg.de,
suto@phys.s.u-tokyo.ac.jp, evrard@umich.edu}

\received{2001 July 11}
\accepted{2001 September 20}

\begin{abstract}
We develop a phenomenological model to predict the clustering of dark
matter halos on the light-cone by combining several existing theoretical
models.  Assuming that the velocity field of halos on large scales is
approximated by linear theory, we propose an empirical prescription of a
scale-, mass-, and time-dependence of halo biasing.  We test our model
against the Hubble Volume $N$-body simulation and examine its validity
and limitations.  We find a good agreement in two-point correlation
functions of dark matter halos between the phenomenological model
predictions and measurements from the simulation for $R>5h^{-1}$Mpc both
in the real and redshift spaces.  Although calibrated on the mass scale
of groups and clusters and for redshifts up to $z\sim2$, the model is
quite general and can be applied to a wider range of astrophysical
objects, such as galaxies and quasars, if the relation between dark
halos and visible objects is specified.
\end{abstract}


\altaffiltext{1}{Present address: 
National Astronomical Observatory, Mitaka 181-8588, Japan}
\altaffiltext{2}{Max-Planck-Institut f\"ur Astrophysik,
Karl-Schwarzschild-Strasse 1,  85748 Garching, Germany}
\altaffiltext{3}{Department of Physics and Research 
Center for the Early Universe
(RESCEU),  School of Science, University of Tokyo, Tokyo 113-0033,
Japan.}
\altaffiltext{4}{Departments of Physics and Astronomy, 
University of Michigan, Ann Arbor, MI 48109-1120}

\section{Introduction}

Clustering properties of luminous objects such as galaxies, clusters of
galaxies and QSOs are useful tools not only in studying the nature of
those objects but also in probing the cosmology.  Current popular models
predict that the cosmic structures evolved
by gravitational instability from primordial fluctuations of mass
density field generated through an inflationary epoch.  The strongest
support for this picture comes from recent detections of multiple peaks in the
angular power spectrum of the cosmic microwave background radiation by
the Boomerang \citep{boomerang01} and MAXIMA-I \citep{maxima01}
experiments.  On the other hand, our knowledge about cosmic structures
after the last scattering epoch, especially at high redshifts $z>1$, is
relatively poor, mostly because of observational costs associated with
mapping the structure of many distant faint objects.  Thanks to recent
developments in instrument technology, this situation is improving
dramatically.  Large flows of data from the on-going wide-field galaxy
and QSO redshift surveys, e.g., the Two-degree field (2dF) and the Sloan
Digital Sky Surveys (SDSS), promise a new era of {\it precision
cosmology}.

How accurately can we understand the nature of clustering of objects
that will be precisely measured by these on-going surveys ?  To
construct a theoretical model of clustering of visible objects (galaxy,
cluster of galaxies and QSOs) is not simple because it requires a
detailed understanding of the biasing relation between those objects and
the distribution of underlying dark mass.  Popular models of the biasing
based on the peak \citep{Kaiser84,bbks86} or the Press-Schechter theory
\citep{mw96} are successful in capturing some essential features of
biasing \citep{js98}.  None of the existing models of bias, however,
seems to be sophisticated enough for the coming precision cosmology era.
Development of a more detailed theoretical model of bias is needed.

One way to understanding the clustering of objects is to describe it in
terms of {\it dark matter halos}.  The standard picture of structure
formation predicts that the luminous objects form in a gravitational
potential of dark matter halos.  Therefore, a detailed description of
halo clustering is the most basic step toward understanding the
clustering of those objects. 
Eventually the halo model can be combined with the
relation between the halos and luminous objects which has been
separately investigated numerically and/or (semi-)analytically, e.g.,
\citet{KH00}.

The purpose of this paper is to improve theoretical predictions for
clustering of luminous objects in large observational catalogs by
developing a theoretical model of clustering of {\it dark matter halos}
expected along the past light-cone of an observer.  A special attention
is payed to the scale-, time-, and mass-dependence of halo biasing.  To
do this, we combine several existing theoretical models including
nonlinear gravitational evolution, the peculiar velocities of halos, and
halos biasing.  We also include the light-cone effect which is crucial
when one analyzes data distributing over a broad redshift range.  We
then test the resulting predictions directly against a light-cone output
from large $N$-body simulations.  This work presents a natural
generalization of our previous paper (Hamana, Colombi, \& Suto 2001;
hereafter HCS) discussing the clustering of dark matter on the light
cone. Nevertheless this line of research benefits greatly from the
modeling huge spatial volumes in simulations, a situation which has
become possible only recently.

\section{Light-cone output and snapshot data from the Hubble volume
simulation}

In the following analysis, we use both ``light-cone output'' and
snapshot data produced from the Hubble Volume $\Lambda$CDM simulation
\citep{evrard01}.  The initial CDM power spectrum is computed by CMBFAST
\citep{seljak96} assuming that $\Omega_{\rm b}=0.04$ and $\Omega_{\rm
CDM}=0.26$, and is normalized so that $\sigma_{8}=0.9$.  The background
cosmology is spatially-flat with matter density $\Omega_{\rm
m}=\Omega_{\rm CDM}+\Omega_{\rm b}=0.3$, cosmological constant
$\Omega_{\Lambda}=0.7$ and the Hubble constant $H_{0}=70$
km$\cdot$s$^{-1}\cdot$Mpc$^{-1}$.  The simulation employs $N=10^9$ dark
matter particles in a box of length 3000$h^{-1}$Mpc on a side.  The mass
per particle is $2.25\times 10^{12} h^{-1}M_{\odot}$.  The light-cone
output is generated in the following manner; we locate a fiducial
observer at a corner of the simulation box at $z=0$.  The position and
velocity of each particle are recorded whenever it crosses the past
light-cone of this observer, and these coordinates are accumulated in a
single data file.  We use the ``deep wedge'' output\footnote{For details
of the output formats of the Hubble volume simulation, see the Virgo
archive web site http://www.mpa-garching.mpg.de/Virgo/hubble.html} which
subtends a 81.45 square-degree field directed along a diagonal of the
simulation box up to $z=4.4$.  These data automatically include the
evolution of clustering with look-back time (distance from the
observer), which is essential in comparing models and observations of
objects distributed over a broad range of redshift.

We identify dark matter halos on the light-cone using the standard
friends-of-friends algorithm with a linking parameter of $b=0.164$ (in
units of the mean particle separation).  \citet{jenkins01} show that
such an algorithm produces a set of clusters whose mass function is well
fit by a single functional form. We set the minimum mass of the halos as
$2.2\times 10^{13} h^{-1}M_{\odot}$, which consists of 10 simulation
particles.  
 
We find that the Press-Schechter model underpredicts the cumulative 
mass function of our halos with  $M>2.2\times 10^{13} h^{-1}M_{\odot}$ at
$z>1$, while Sheth \& Tormen (1999, hereafter ST99) overpredicts beyond
$z\sim1.5$. This tendency is consistent with the previous finding of
\citet{jenkins01} that ST99 overestimates the number of halos when
ln($\sigma^{-1}$) becomes large.

In \S 3.2 we also use the halo catalogue identified in the $z=0$
snapshot data of the Hubble volume simulation to study the mass- and
scale-dependence of halo biasing in detail.  The halos are identified in
the same manner as described above except that the minimum halo mass is
$6.8\times 10^{13} h^{-1}M_{\odot}$ (30 simulation particles).  The
total number of the identified halos in the $(3000h^{-1}$Mpc$)^3$-cube
is 1,560,995. We note that the mass function and clustering of this this
halo catalogue at $z=0$ were already studied by \citep{jenkins01}, and
by \citep{Colberg00}, respectively.  Our analysis below aims at a
detailed modeling of the halo biasing properties at $z=0$ in order to
calibrate the empirical halo biasing model on the light-cone.

\section{Statistics of halos on the light-cone}

\subsection{Theoretical predictions of two-point correlation functions
on the light-cone}

As emphasized by \citet{suto99}, for instance, observations of
high-redshift objects are carried out only through the past light-cone
defined at $z=0$, and the corresponding theoretical modeling should
properly take account of relevant physical effects.  Those include (i)
nonlinear gravitational evolution, (ii) linear and nonlinear redshift
space distortion, (iii) selection function of the target objects, and
(iv) scale-, mass- and time-dependent biasing of those objects.  In the
present section we describe a model for the two-point statistics for
dark matter halos with {\it all} the above effects properly considered.
In what follows, we briefly describe the outline of our modeling (see
HCS for details) focusing on those issues specific to dark matter
halos.

Gravitational evolution of mass fluctuations can be accurately modeled
by adopting a fitting formula of \citet{pd96} for the nonlinear power
spectrum in real space, $P^{\rm R}_{\rm PD}(k,z)$.  Then the nonlinear
power spectrum {\it in redshift space} is given as \citep{Kaiser87,pd96}:
\begin{equation}
P^{\rm S}(k,\mu,z)=P^{\rm R}_{\scriptscriptstyle\rm PD}(k,z)
[1+\beta_{\rm halo}\mu^{2}]^{2}
D_{\rm vel}[k\mu\sigma_{\rm halo}],
\label{nonlinear}
\end{equation}
where $\mu$ is the direction cosine in $k$-space, $\sigma_{\rm halo}$ is
 the one-dimensional {\it pair-wise} velocity dispersion of halos, and
$\beta_{\rm halo} \equiv f(z)/b_{\rm halo}$.
In the above expression, $f(z)$ is the logarithmic derivative of the
linear growth rate $D(z)$ with respect to the scale factor, and $b_{\rm
halo}$ is the halo bias factor.  While both $\sigma_{\rm halo}$ and
$b_{\rm halo}$ depend on the halo mass $M$, separation $R$, and $z$ in
reality, we neglect their scale-dependence in computing the redshift
distortion, and adopt the halo number-weighted averages:
\begin{eqnarray}
\label{eq:sigma_halo}
\sigma_{\rm halo}^2(>M,z) &\equiv& 
 \frac{\displaystyle \int_M^\infty 2D^2(z)\sigma^2(M,z=0)\, 
   n_{\scriptscriptstyle\rm J}(M,z) dM}
 {\displaystyle\int_M^\infty n_{\scriptscriptstyle\rm J}(M,z) dM} , \\
\label{eq:b_halo}
b_{\rm halo}(>M,z) &\equiv& 
 \frac{\displaystyle\int_M^\infty b_{\scriptscriptstyle\rm ST}(M,z)\, 
  n_{\scriptscriptstyle\rm J}(M,z) dM}
 {\displaystyle\int_M^\infty n_{\scriptscriptstyle\rm J}(M,z) dM} , 
\end{eqnarray}
where we adopt the halo mass function $n_{\scriptscriptstyle\rm J}(M,z)$
 fitted by \citet{jenkins01} and the mass-dependent halo bias factor
 $b_{\scriptscriptstyle\rm ST}(M,z)$ proposed by ST99.  The value of
 $\sigma(M,z=0)$, the halo center-of-mass velocity dispersion at $z=0$,
 is modeled following Yoshida, Sheth \& Diaferio (2001):
\begin{equation}
\sigma(M,z=0) = \frac{430/\sqrt{3}}
{1+(M/2.487\times10^{16}h^{-1}M_\odot)^{0.284}}
  \;\mbox{km s}^{-1}.
\label{sheth-diaferio}
\end{equation}
In deriving equation (\ref{eq:sigma_halo}), we follow \citet{sd01} and
assume that the evolution of halo velocities is well approximated by 
linear theory, and in order to relate the halo velocity dispersion to
the pairwise velocity dispersion, we have assumed no velocity
correlation between halos.  We use the Lorentzian damping function
$D_{\rm vel}$ in $k$-space for mass \citep{magira00}, and the Gaussian
for halos \citep{uis93,ysd01}.

\subsection{Modeling scale and mass dependence of the halo biasing}

The most important ingredient in describing the clustering of halos is
their biasing properties. The mass-dependent halo bias model was
developed by \citet{mw96} based on the extended Press-Schechter theory
\citep{lc93}. This biasing model is improved empirically by
\citet{Jing98} and ST99 so as to more accurately reproduce the
mass-dependence in $N$-body simulation results. We further attempt to
incorporate the scale-dependence on the basis of the results of Taruya
\& Suto (2000, hereafter TS00) in which the scale-dependence arises as a
natural consequence of the formation epoch distribution of
halos. \citet{yoshikawa01} exhibit that the scale-dependence of the TS00
model agrees with their numerical simulations although the amplitude is
larger because the volume exclusion effect is not properly taken into
account in the model. Therefore we construct an empirical halo bias
model of the two-point statistics which reproduces the scale-dependence
of the TS00 bias with the amplitude fixed by the mass-dependent ST99
bias on linear scales.  We find that the required behavior is well
described by the following simple fitting formula:
\begin{equation}
\label{eq:bhalo_MRz}
b_{\rm halo}(M,R,z)=
  b_{\scriptscriptstyle\rm ST}(M,z) 
\left[ 1.0+ b_{\scriptscriptstyle\rm ST}(M,z)\sigma_R(R,z) \right]^{0.15}
\end{equation}
for $R>2R_{\rm vir}(M,z)$, and otherwise 0, where $R_{\rm vir}(M,z)$ is
the virial radius of the halo of mass $M$ at $z$ and $\sigma_R (R,z)$ is
the mass variance smoothed over the top-hat radius $R$.  While the above
cutoff below $2R_{\rm vir}(M,z)$ is simply intended to incorporate the
halo exclusion effect very roughly, we find it a reasonable
approximation as will be shown below.  

\vspace{0.5cm}
\centerline{{\vbox{\epsfxsize=8.6cm\epsfbox{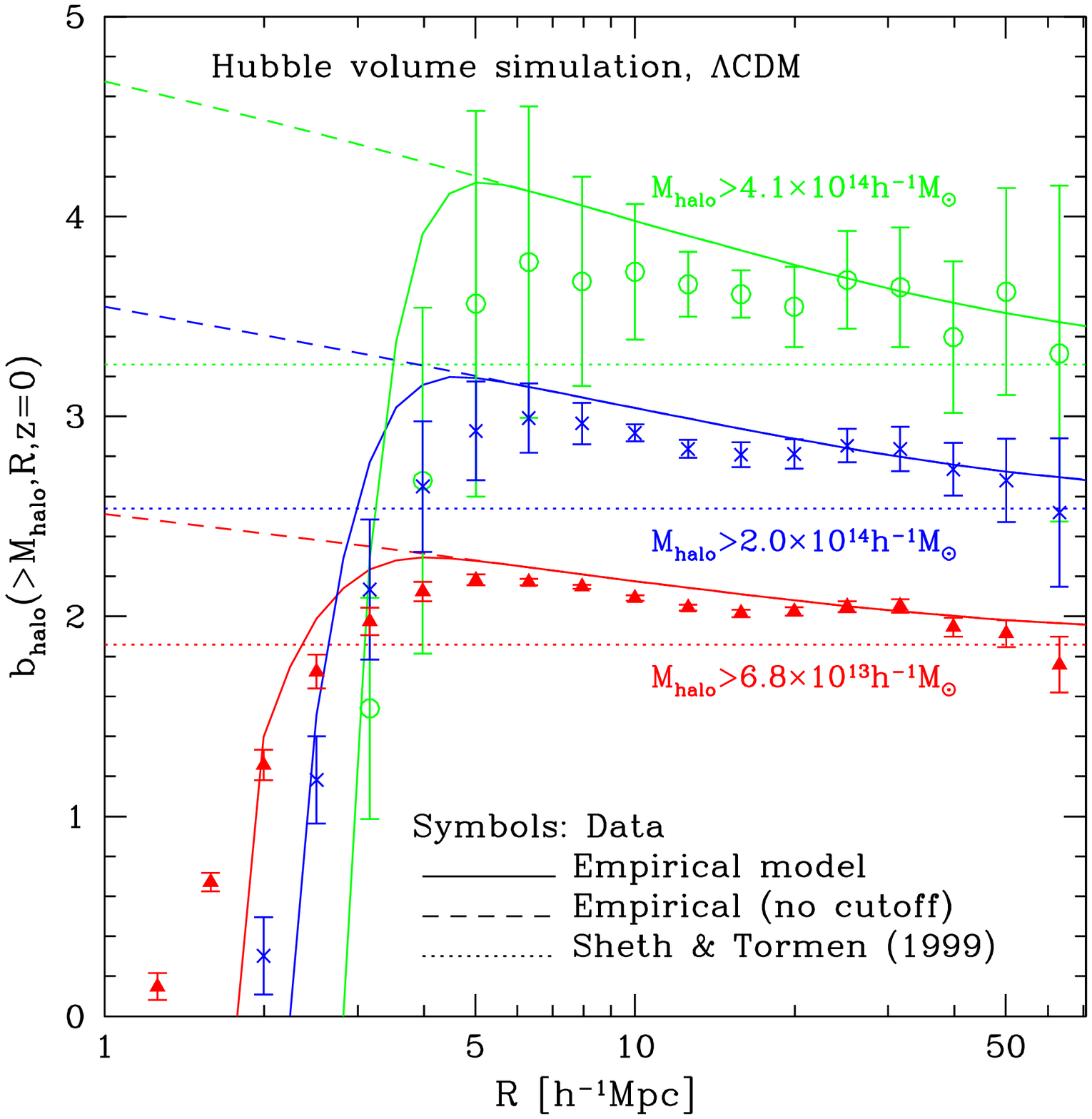}}}}
\figcaption{Halo biasing parameter defined by the two-point correlation
 functions on the constant-time hyper-surface in real space ($z=0$);
 solid and dotted lines indicate our empirical model
 [eq.~(\ref{eq:bhalo_MRz})], and the scale-independent model by ST99,
 respectively. The dashed lines correspond to our model without
 correction for the halo exclusion effect. \label{fig:bias_r_m}}  
\vspace{0.5cm}

We test this empirical bias model against the halo catalogue generated
from the snapshot data at $z=0$.  To do this, we compute the two-point
correlation functions of halos of mass $M_{\rm halo}>M_{\rm min}$ in
real space, then we divide them directly by the corresponding mass
correlation function.  We adopt the estimator $\xi=(DD-2DR+RR)/RR$
\citep{ls93} with the standard bootstrap method with 200 random
re-samplings.  Figure~\ref{fig:bias_r_m} compares the resulting bias
factor of halos, $b_{\rm halo}(>M,R,z=0)$; open circles, crosses, and
filled triangles for $M_{\rm min}=4.1\times10^{14}h^{-1}M_\odot$,
$2.0\times10^{14}h^{-1}M_\odot$, and $6.8\times10^{13}h^{-1}M_\odot$,
respectively.  The dotted horizontal lines indicate the
$b_{\scriptscriptstyle\rm ST}(M,z)$, and our model predictions
(eq.[\ref{eq:bhalo_MRz}]) are plotted in solid lines. Given the simple
formula that we adopted, the agreement with the numerical simulations at
$z=0$ is satisfactory.

Then our empirical halo bias model can be applied to the two-point
correlation function of halos at $z$ in redshift space as
\begin{equation}
\label{eq:xihalo_z}
\xi_{\rm halo}(M,R,z) = b^2_{\rm halo}(M,R,z)
\int_0^\infty P^{\rm S}(k,z) \frac{\sin kR}{kR} 
 \frac{k^2 dk}{2\pi^2} .
\end{equation}
Finally, the correlation function of halos on the light-cone is computed
by averaging over the appropriate halo number density and the comoving
volume element between the survey range $z_{\rm min}<z<z_{\rm max}$
\citep{Matarrese97,Moscardini98,ys99,suto00}:
\begin{eqnarray}
\label{eq:xihalolc}
    \xi_{\rm halo}^{\rm\scriptscriptstyle {LC}}(>M,R) 
&=& {
   {\displaystyle \int_M^\infty dM 
     \int_{z_{\rm min}}^{z_{\rm max}} dz 
     {dV_{\rm c} \over dz} ~n_{\scriptscriptstyle\rm J}^2(M,z)
    \xi_{\rm halo}(M,R,z)
    }
\over
    {\displaystyle \int_M^\infty dM 
     \int_{z_{\rm min}}^{z_{\rm max}} dz 
     {dV_{\rm c} \over dz}  ~n_{\scriptscriptstyle\rm J}^2(M,z)
     }
} ,
\end{eqnarray}
where $dV_{\rm c}/dz$ is the comoving volume element per unit solid
angle (HCS).  Although our modeling is not completely self-consistent
in the sense that the scale-dependence of the halo biasing factor is
neglected in describing the redshift distortion (\S 3.2), the above
prescription is supposed to provide a good approximation since the
scale-dependence in the biasing is of secondary importance in the
redshift distortion effect of halos.

\subsection{Clustering on the light-cone}

\begin{figure*}[tbp]
\begin{center}
\leavevmode
\epsfxsize=14.0cm 
\epsfbox{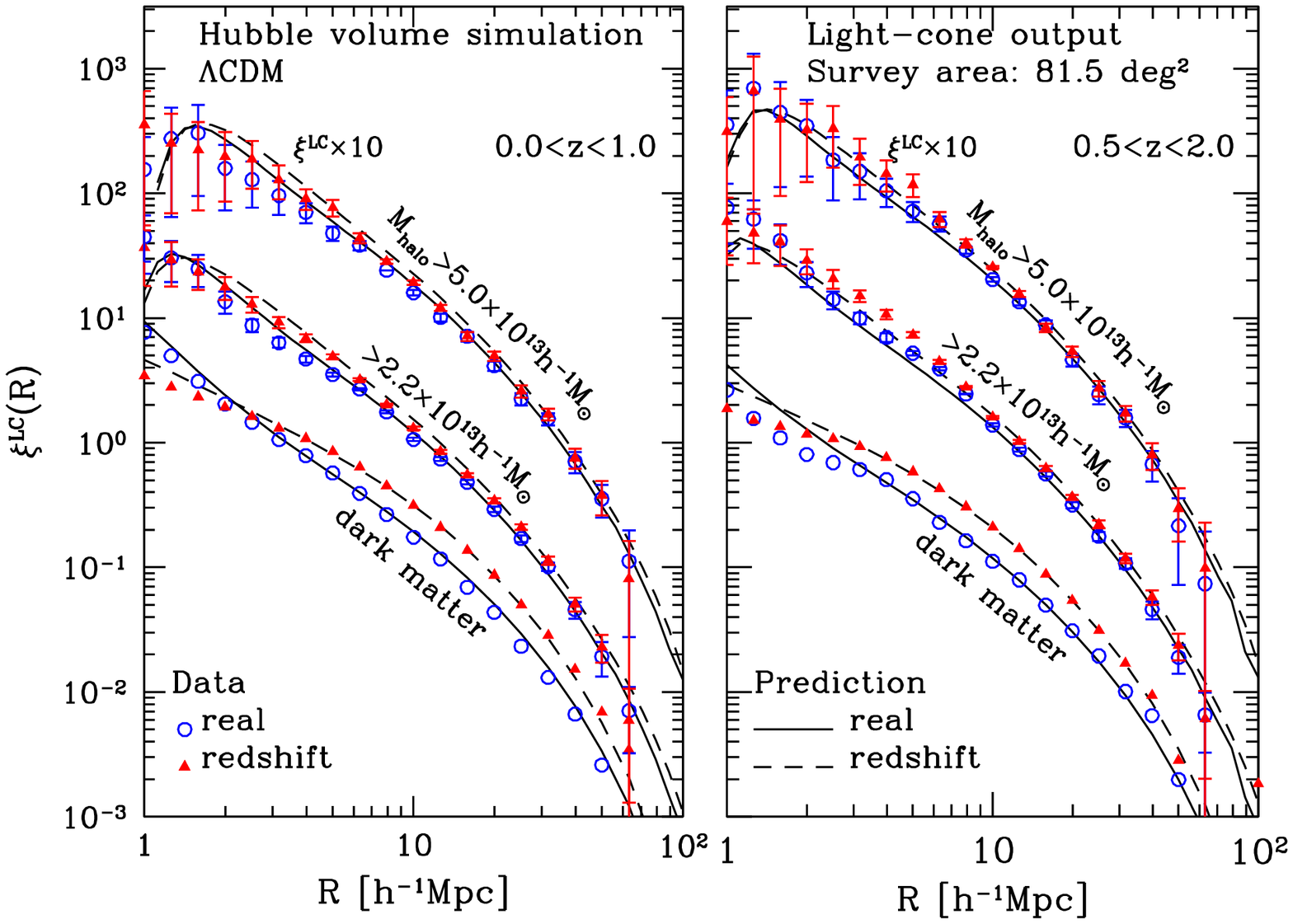}
\caption{Two-point correlation functions of halos on the light-cone;
simulation results (symbols; open circles and filled triangles for real
and redshift spaces, respectively) and our predictions (solid and dotted
lines for real and redshift spaces, respectively).  The error bars
denote the standard deviation computed from 200 random re-samplings of
the bootstrap method.  Upper set is for halos with $M_{\rm halo}\ge
5.0\times 10^{13}h^{-1} M_\odot$, middle set is for $M_{\rm halo}\ge
2.2\times 10^{13}h^{-1} M_\odot$, and lower set is for the dark matter.
Notice that the amplitude of the upper sets is increased by an order of
magnitude for clarity. \label{fig:haloxilc}}
\end{center}
\end{figure*}

The two-point correlation functions on the light-cone are plotted in
Figure~\ref{fig:haloxilc} for halos with
$M>5.0\times10^{13}h^{-1}M_\odot$, $M>2.2\times10^{13}h^{-1}M_\odot$ and
dark matter from top to bottom. The range of redshift is $0<z<1$ ({\it
Left panel}) and $0.5<z<2$ ({\it Right panel}). Predictions in redshift
and real spaces are plotted in dashed and solid lines, while simulation
data in redshift and real spaces are shown in filled triangles and open
circles, respectively.

Our model and simulation data also show quite good agreement for dark
halos at scales larger than $5h^{-1}$Mpc. Below that scale, they start
to deviate slightly in a complicated fashion depending on the mass of
halo and the redshift range.  This discrepancy may be ascribed to both
the numerical limitations of the current simulations and our rather
simplified model for the halo biasing (eq.[\ref{eq:bhalo_MRz}]).
Nevertheless the clustering of {\it clusters} on scales below
$5h^{-1}$Mpc is difficult to determine observationally anyway, and our
model predictions differ from the simulation data only by $\sim 20$
percent at most. Therefore we conclude that in practice our empirical
model provides a successful description of halo clustering on the
light-cone.

\section{Conclusions and discussion}

We develop a phenomenological model to predict the clustering of dark
matter halos on the light-cone by combining several existing theoretical
models.  Combining the TS00 bias model with the ST99 mass function
model, we are, for the first time, able to construct a halo biasing
model that reproduces well the mass- and radial-dependence measured in
the Hubble Volume simulation output data.  Once calibrated with the
$z=0$ snapshot data, we find that our model agrees well with the
two-point correlation functions of the simulated halos up to $z=2$ in
both real and redshift spaces.  Although we show that this
phenomenological model of halo clustering provides accurate predictions
for the two-point correlation function of halos over a limited range in
mass and redshift, we anticipate that it can be applied to a wider range
of scales.  This opens up application to modeling observations of
various astrophysical objects, such as galaxies, clusters of galaxies
and quasars, under model-specific assumptions for the relation between
dark halos and luminous objects.

\acknowledgments

T.H. thanks G.B\"orner and M.Bartlemann for the hospitality during his
stay at MPA where most of the present work is performed. He also
acknowledges support from Research Fellowships of the Japan Society for
the Promotion of Science.  The simulation used here was carried out by
the Virgo Consortium and the light-cone data are publicly available at
http://www.mpa-garching.mpg.de/Virgo. This research was supported in
part by the Grant-in-Aid by the Ministry of Education, Science, Sports
and Culture of Japan (07CE2002, 12640231).  A.E.E. acknowledges support
from NSF grant AST-9803199, NASA grant NAG5-7108 and the Scientific
Visitor Program of the Carnegie Observatories in Pasadena.

\bigskip
\bigskip



\end{document}